\documentclass{article}

\usepackage{graphicx}
\graphicspath{{converted_graphics/}}
\begin{document}

\begin{center}
{\LARGE Physically Derived Rules for Simulating}\vskip 6pt

{\LARGE Faceted Crystal Growth using Cellular Automata}\vskip16pt

{\Large Kenneth G. Libbrecht}\footnote{e-mail address: kgl@caltech.edu; for updates on this and other papers, see http://www.its.caltech.edu/~atomic/publist/kglpub.htm
}\vskip4pt

{\large Department of Physics, California Institute of Technology}\vskip-1pt

{\large Pasadena, California 91125}\vskip-1pt

\vskip18pt

\hrule \vskip1pt \hrule
\vskip 14pt
\end{center}

\noindent \textbf{Abstract. We derive a set of algorithms for simulating the
diffusion-limited growth of faceted crystals using local cellular automata.
This technique has been shown to work well in reproducing realistic crystal
morphologies, and the present work provides a more rigorous physical
foundation that connects the numerical code to the physics of attachment
kinetics and diffusion dynamics. We then apply these algorithms to examine a
novel morphological transition in the growth of thin plate-like crystals. }

\section{\noindent Introduction}

The formation of complex structures during solidification often results from
a subtle interplay of nonequilibrium, nonlinear processes, for which
seemingly small changes in molecular dynamics at the nanoscale can produce
profound morphological changes at all scales. One popular example of this
phenomenon is the formation of snow crystals, which are ice crystals that
grow from water vapor in an inert background gas. Although this is a
relatively simple, monomolecular system, snow crystals display a remarkable
variety of columnar and plate-like forms, and much of the phenomenology of
their growth remains poorly understood, even at a qualitative level \cite%
{libbrechtreview}.

Viewed broadly, snow crystal structures result from diffusion-limited
crystal growth in the presence of strongly anisotropic attachment kinetics,
and similar circumstances occur in a large class of solidification problems.
We can break such problems down into two main physical components: 1) the
attachment kinetics that describes the molecular growth dynamics at the
solid surface, and 2) particle transport via diffusion to the growing
crystal. In the present paper we focus on the latter problem -- numerically
solving the diffusion equation to model faceted crystal growth.

Computational models of diffusion-limited solidification have typically been
divided into two broad camps -- `front-tracking' models, in which one keeps
track of the solidification interface explicitly (e.g. \cite{yokoyama,
yokoyamak, kglcyl}), and `phase-field' models, in which the solidification
front is numerically smoothed and not explicitly tracked (e.g. \cite{104,
morphology}). To date, both techniques have had considerable success in
modeling simple dendritic growth, but neither has been able to
satisfactorily model complex morphological structures in the presence of
strong faceting \cite{108, 109, phase, polygon}, owing to the appearance of
dynamical and numerical instabilities.

A third computation technique - local cellular automata (LCA) - has recently
been applied to the problem of modeling faceted crystal growth with
excellent initial success, and LCA models have produced realistic-looking
snow crystal growth simulations in both 2D \cite{reiter, cellular} and 3D 
\cite{gravnergriffeath}. To date, however these local lattice models have
incorporated largely \textit{ad hoc }growth rules (albeit physically
motivated to some extent), so the connection between the resulting
simulations and real crystal growth remains tenuous. These algorithms
yielded structures that bear a striking resemblance to many features seen in
real snow crystals, but the similarities are largely qualitative. As a
result, these models cannot be directly compared with crystal growth
measurements to provide quantitative insights into growth dynamics.

The goal of this paper is to produce more accurate, physically derived rules
for modeling with cellular automata, thus providing a direct connection
between the numerical code and the physics of attachment kinetics and
diffusion dynamics governing faceted crystal growth. The algorithms derived
below provide a more rigorous physical foundation for using LCA techniques
in crystal growth models. After deriving growth algorithms below for the
case of snow crystal formation, we then examine the properties of the LCA
model and describe a novel morphological transition seen in plate-like
growth.

\section{Crystal Growth Dynamics}

We focus our discussion on the growth of snow crystals, as this is perhaps
the most studied example of strongly faceted, diffusion-limited crystal
growth. Having such a focus allows a more direct look at the relevant
physics, which becomes somewhat obscured in a more general discussion.
Extending the algorithms to other materials should be relatively
straightforward, provided the growth dynamics are of a similar character.

We begin by assuming the "standard model" of snow crystal growth \cite%
{libbrechtreview}, for which we can first write the growth velocity normal
to the surface in terms of the Hertz-Knudsen formula 
\begin{eqnarray}
v_{n} &=&\alpha \frac{c_{sat}}{c_{solid}}\sqrt{\frac{kT}{2\pi m}}\sigma
_{surf}  \label{kinetic} \\
&=&\alpha v_{kin}\sigma _{surf}  \nonumber
\end{eqnarray}%
where the latter expression defines the velocity $v_{kin}.$ In this
expression $kT$ is Boltzmann's constant times temperature, $m$ is the mass
of a water molecule, $c_{solid}=\rho _{ice}/m$ is the number density for
ice, $\sigma _{surf}=(c_{surf}-c_{sat})/c_{sat}$ is the supersaturation just
above the growing surface, $c_{surf}$ is the water vapor number density at
the surface, and $c_{sat}(T)$ is the equilibrium number density above a flat
ice surface. The dimensionless parameter $\alpha $ is known as the \textit{%
condensation coefficient}, and it embodies the surface physics that governs
how water molecules are incorporated into the ice lattice, collectively
known as the \textit{attachment kinetics}.

The attachment kinetics can be nontrivial, so in general $\alpha $ will
depend on $T$, $\sigma _{surf},$ and perhaps on the surface structure and
geometry, surface chemistry, and other factors. If molecules striking the
surface are instantly incorporated into it, then $\alpha =1;$ otherwise we
must have $\alpha \leq 1.$ The appearance of crystal facets indicates that
the growth is limited in part by attachment kinetics, so we must have $%
\alpha <1$ on faceted surfaces.

This model assumes that the attachment kinetics are purely local, without
significant large-scale surface diffusion, especially around corners between
facets. It also assumes that all complex aspects of the molecular dynamics
at the ice surface, which may include surface melting and any number of
other details, can be absorbed into some parameterization of $\alpha $ as a
function of external parameters. There is currently no evidence that this
"standard model" is incorrect \cite{libbrechtreview}, but at the same time
we cannot prove that the attachment kinetics is always well described by
Equation \ref{kinetic}. For the present discussion, we will assume that this
expression is valid for circumstances relevant to snow crystal growth.

Particle transport through the air surrounding a growing crystal is
described by the diffusion equation 
\begin{equation}
\frac{\partial c}{\partial t}=D\nabla ^{2}c
\end{equation}%
where $c(x)$ is the water molecule number density surrounding the crystal
and $D$ is the diffusion constant. The timescale for diffusion to adjust the
vapor concentration in the vicinity of a crystal is $\tau
_{diffusion}\approx R^{2}/D,$ where $R$ is a characteristic crystal size.
This is to be compared with the growth time, $\tau _{growth}\approx
2R/v_{n}, $ where $v_{n}$ is the growth velocity of the solidification front
normal to the surface. The ratio of these two timescales is the Peclet
number, $p=Rv_{n}/2D.$ For typical growth rates of snow crystals we find $%
p< 10^{-5},$ which means that diffusion adjusts the particle density
around the crystal much faster than the crystal shape changes. In this case
the diffusion equation reduces to Laplace's equation, $\nabla ^{2}c=0,$
which must be solved with the appropriate boundary conditions. Using this
slow-growth limit often simplifies the problem considerably in comparison to
much of the literature on diffusion-limited solidification \cite{libbrecht11}%
.

The continuity equation at the interface gives 
\begin{equation}
v_{n}=\frac{D}{c_{solid}}\left( \widehat{n}\cdot \overrightarrow{\nabla }%
c\right) _{surf}=\frac{c_{sat}D}{c_{solid}}\left( \widehat{n}\cdot 
\overrightarrow{\nabla }\sigma \right) _{surf}
\end{equation}%
where $\sigma (x)=[c(x)-c_{sat}]/c_{sat}$ and we are assuming the isothermal
case, so $c_{sat}$ is independent of spatial position. The latter assumption
means we will be ignoring effects that arise when a growing crystal
experiences a temperature increase associated with the latent heat of
solidification. These effects are generally small and produce results that
are similar to a simple decrease in the supersaturation surrounding the
crystal \cite{libbrechtreview}. Thus we believe we will not be sacrificing
much interesting physics by using the isothermal approximation. We then
write the diffusion equation in terms of the supersaturation field (since $%
c_{sat}$ is constant) as%
\begin{equation}
\frac{\partial \sigma }{\partial t}=D\nabla ^{2}\sigma
\end{equation}

The attachment coefficient $\alpha $ is not well known for ice, but on facet
surfaces it appears to be well described by nucleation-limited growth \cite%
{libbrechtreview, libbrechtdata, libbrechtnew}, which gives%
\begin{equation}
\alpha (\sigma )\approx A\exp \left( -\sigma _{0}/\sigma \right)
\end{equation}%
where $A$ and $\sigma _{0}$ are parameters that depend on temperature and
are different for the basal and prism facets. For growth governed by spiral
dislocations, we expect \cite{saitobook}%
\begin{equation}
\alpha (\sigma )\approx C\sigma
\end{equation}%
where $C$ is independent of $\sigma $ (but depends on external parameters
such as temperature), which gives $v_{n}\sim \sigma ^{2}.$

\section{Derivation of Growth Algorithms}

\subsection{The 1D Problem}

Because of its simplicity, it is instructive to begin with the 1D crystal
growth problem, for which we assume a linear array of grid elements, or
pixels, of width $\Delta x.$ Each pixel is assumed to be filled with either
ice or air at any given time, and each air pixel has a well-defined
supersaturation $\sigma (x,t)$ associated with it. The discrete form of the
1D diffusion equation is%
\begin{eqnarray}
\frac{\partial \sigma }{\partial t} &=&D\frac{d^{2}\sigma }{dx^{2}} \\
\frac{\sigma (x,t+dt)-\sigma (x,t)}{dt} &=&D\left[ \frac{\sigma
(x+dx)-2\sigma (x)+\sigma (x-dx)}{dx^{2}}\right]  \nonumber
\end{eqnarray}%
which yields%
\begin{equation}
\sigma (x,t+\Delta t)=\Delta \tau \sigma (x-\Delta x,t)+\left( 1-2\Delta
\tau \right) \sigma (x,t)+\Delta \tau \sigma (x+\Delta x,t)  \label{relax}
\end{equation}%
where $\Delta t$ is the physical time step in our simulation code and $%
\Delta x$ is the grid spacing (assumed uniform). The quantity%
\begin{equation}
\Delta \tau =\frac{D\Delta t}{\left( \Delta x\right) ^{2}}
\end{equation}%
is a dimensionless time-step parameter. We note that $\left( \Delta x\right)
^{2}/D$ is one natural time scale of the problem, equal the time required
for diffusion to adjust the supersaturation field over a distance $\Delta x$%
. We define $\tau =\sum \Delta \tau $ to be the dimensionless physical time
in our problem.

We use Equation \ref{relax} to effectively relax the supersaturation field
to a suitable solution of the diffusion equation. Choosing $\Delta \tau $
too small will require an excessive amount of computer time, while taking $%
\Delta \tau >1$ leads to instabilities near sharp edges (\textit{e.g.}, at
physical boundaries) in $\sigma $. It appears that $\Delta \tau =0.5$ is
close to optimal, yielding the simple algorithm%
\begin{equation}
\sigma (x,\tau +\Delta \tau )=\frac{1}{2}\sigma (x-\Delta x,\tau )+\frac{1}{2%
}\sigma (x+\Delta x,\tau )
\end{equation}%
for the 1D case. This expression is used to propagate $\sigma $ in time for
regions away from the growing crystal surface, where diffusion alone affects
the particle dynamics.

Our model crystal grows as water molecules diffuse in from the outer
boundaries of the system and attach to the ice surface. For boundary pixels
(i.e., air pixels that are adjacent to ice pixels), we see physically that
the ice surface effectively "drains" the supersaturation as water vapor
condenses, which in turn causes the solidification front to move with a
velocity $v=\alpha v_{kin}\sigma .$ This "draining" reduces the
supersaturation of boundary pixels according to%
\begin{equation}
\sigma (\tau +\Delta \tau )=\sigma (1-\alpha \Delta \xi \Delta \tau )
\label{changingsigma}
\end{equation}%
where $\Delta \xi =\Delta x/X_{0}$ is a dimensionless pixel size in the
problem and%
\begin{eqnarray}
X_{0} &=&\frac{c_{sat}}{c_{solid}}\frac{D}{v_{kin}}  \label{X0} \\
&=&D\sqrt{\frac{2\pi m}{kT}}  \nonumber \\
&\approx &\left( \frac{D}{D_{air}}\right) (0.15\ \mu m) 
\nonumber
\end{eqnarray}%
is the natural physical scale of the problem. Here $D_{air}=2\times 10^{-5}$
m/sec$^{2}$ is the diffusion constant for air at one atmospheric pressure,
and the numerical value pertains to ice crystal growth.

We can combine Equation \ref{changingsigma} with Equation \ref{relax} to
create a new propagation algorithm for boundary pixels%
\begin{equation}
\sigma (x,\tau +\Delta \tau )=\Delta \tau \sigma _{solid}(\tau )+\left(
1-2\Delta \tau \right) \sigma (x,t)+\Delta \tau \sigma (x+\Delta x,\tau )
\label{d1generalalgorithm}
\end{equation}%
where the boundary pixel is at $x$ and the ice pixel is at $x=x-\Delta x,$
and%
\begin{equation}
\sigma _{solid}(\tau )=\sigma (x,\tau )(1-\alpha \Delta \xi )
\label{sigsolid}
\end{equation}%
With this, the same algorithm can be used for all air pixels, including
boundary pixels, provided one substitutes $\sigma _{solid}$ in the
appropriate place for boundary pixels.

For the conversion of boundary pixels to ice, mass conservation suggests
that we define an accumulated mass parameter $\lambda $ that begins as $%
\lambda =0$ for each air pixel. For boundary pixels, we increment $\lambda $
using%
\begin{equation}
\lambda \rightarrow \lambda +\alpha \sigma \Delta \lambda
\end{equation}%
for each time step, where%
\begin{equation}
\Delta \lambda =\frac{c_{sat}}{c_{solid}}\Delta \tau \Delta \xi
\end{equation}%
When $\lambda \geq 1$, that pixel converts from air to ice.

For ice we have $c_{sat}/c_{solid}\approx 10^{-6},$ while the dimensionless
parameters $\sigma ,$ $\Delta \tau ,$ $\alpha ,$ $\Delta \xi $ will all be
of order unity or perhaps substantially smaller. Thus it would take more
than a million time steps before a boundary pixel becomes an ice pixel. This
situation reflects the small Peclet number in our problem, so the growth is
much slower than the time it takes diffusion to adjust the supersaturation
field. We speed up the code by taking%
\begin{equation}
\Delta \lambda =\Lambda \Delta \tau \Delta \xi  \label{lambda}
\end{equation}%
where $\Lambda $ is an input constant, discussed further below.

\subsection{The 2D problem in Cartesian Coordinates}

Following the 1D analysis, the discrete form of the 2D diffusion equation
gives%
\begin{equation}
\sigma (\tau +\Delta \tau )=\Delta \tau \left[ \sigma (x-\Delta x)+\sigma
(x+\Delta x)+\sigma (y-\Delta y)+\sigma (y+\Delta y)\right] +\left(
1-4\Delta \tau \right) \sigma  \label{d2sigma}
\end{equation}%
where we have dropped indices when doing so does not confuse the expression.
If we take our time parameter to be $\Delta \tau =1/4,$ this becomes%
\begin{equation}
\sigma (\tau +\Delta \tau )=\frac{1}{4}\left[ \sigma (x-\Delta x)+\sigma
(x+\Delta x)+\sigma (y-\Delta y)+\sigma (y+\Delta y)\right]
\end{equation}

For boundary pixels, we again use this expression and substitute%
\begin{equation}
\sigma _{solid}=\sigma (1-\alpha \Delta \xi )
\end{equation}%
for each neighboring ice pixel, as with the 1D case, and we have further
assumed a uniform grid with $\Delta x=\Delta y.$

We again define an accumulated mass parameter $\lambda $ and increment it
with%
\begin{equation}
\lambda \rightarrow \lambda + \sum \alpha \sigma \Delta \lambda
\end{equation}%
for each time step, where the sum is over the number of ice neighbors. When $%
\lambda \geq 1$, that pixel converts from air to ice.

In both these expressions we must choose $\alpha $ with some care, as its
value will depend critically on the number and orientation of neighboring
solid pixels. We label a boundary pixel with $(N_{x},N_{y}),$ where $N_{x}$
is the number of neighboring ice pixels in the $x$ direction and $N_{y}$ is
the number of ice neighbors in the $y$ direction. Both $N_{x}$ and $N_{y}$
can take values 0, 1, or 2, giving nine cases for $(N_{x},N_{y}).$ The cases
are:

(0,0) - the pixel is an air pixel

(0,1) - one ice neighbor in the $y$ direction, so $\alpha =$ $\alpha _{y},$
the physical value appropriate for a $y$ facet surface.

(1,0) - one ice neighbor in the $x$ direction, so $\alpha =\alpha _{x}$ for
an $x$ facet surface.

(1,1) - a kink site, where the growth will not be nucleation-limited, since
the corner provides a source of molecular steps. We do not know $a$ $priori$
what value to use for $\alpha $ on this site, but assume a constant $\alpha
=\alpha _{11}$.

(0,2), (1,2), (2,0), (2,1), (2,2) - these are all unusual cases where the
growth will be fast, so it shouldn't matter much what we choose for $\alpha
, $ as long as it is large.

We can index these possibilities with a single number by computing a
boundary parameter $B=2N_{x}^{2}+N_{y}^{2}$, where $N_{x}$ is the number of $%
x$ neighbors (0, 1, or 2) and $N_{y}$ is the number of $y$ neighbors. We
then have $B=0$ for an air pixel, $B=1$ for a $y$ facet, $B=2$ for an $x$
facet, $B=3$ for a (1,1) kink location, and $B>3$ for all other cases.

If we consider the special case where $\alpha $ is equal to some constant
value, independent of the orientation of the surface with respect to the
crystal lattice, then the growth velocity should equal $v=\alpha
v_{kin}\sigma $ for all surfaces. For the (01) or (10) facet surfaces 
\footnote{%
The notation here -- integers in parentheses without commas -- refers to the
usual Miller indices defining specific crystal surfaces. This should not be
confused with our previous notation -- integers in parentheses with commas
-- which we used to label boundary pixels.} in this constant-$\alpha $ case,
we take $\alpha _{x}=\alpha _{y}=\alpha $, while an analysis of the growth
of a (11) surface shows that we must take $\alpha _{11}=\alpha /\sqrt{2}$ if
the above algorithm is to produce the correct growth velocity.

\subsection{The 2D Problem in Cylindrical Coordinates}

The discrete version of the Laplacian in cylindrical coordinates yields%
\begin{equation}
\sigma (\tau +\Delta \tau )=\Delta \tau \left[ \left( 1-\frac{\Delta r}{2r}%
\right) \sigma (r-\Delta r)+\left( 1+\frac{\Delta r}{2r}\right) \sigma
(r+\Delta r)+\sigma (z-\Delta z)+\sigma (z+\Delta z)\right] +\left(
1-4\Delta \tau \right) \sigma
\end{equation}%
where we have assumed $\partial \sigma /\partial \theta =0$ to reduce the
problem to 2D, thus yielding cylindrically symmetric crystal structures. The
problem then proceeds essentially as with the 2D case in rectangular
coordinates, and similar $(1-\Delta r/2r)$ correction factors are needed for 
$\sigma _{solid}$ and $\Delta \lambda .$ With the preferred value of $\Delta
\tau =1/4$ we have%
\begin{equation}
\sigma (\tau +\Delta \tau )=\frac{1}{4}\left[ \left( 1-\frac{\Delta r}{2r}%
\right) \sigma (r-\Delta r)+\left( 1+\frac{\Delta r}{2r}\right) \sigma
(r+\Delta r)+\sigma (z-\Delta z)+\sigma (z+\Delta z)\right]
\end{equation}

In our model, we define the first row of pixels to have $r=0,$ for which we
cannot evaluate the above expression. Instead we revert to the Cartesian
form of the Laplacian to give%
\begin{equation}
\sigma (\tau +\Delta \tau )=\Delta \tau \left[ 4\sigma (r+\Delta r)+\sigma
(z-\Delta z)+\sigma (z+\Delta z)\right] +\left( 1-6\Delta \tau \right) \sigma
\end{equation}%
for those pixels. If we take $\Delta \tau =1/4$ (the same as for $r\neq 0),$
this becomes%
\begin{equation}
\sigma (\tau +\Delta \tau )=\frac{1}{4}\left[ 4\sigma (r+\Delta r)+\sigma
(z-\Delta z)+\sigma (z+\Delta z)\right] -\frac{1}{2}\sigma
\end{equation}%
and the negative term leads to instabilities. We could correct this by
choosing a smaller $\Delta \tau ,$ but this would make the code run more
slowly. An alternative is to use $\Delta \tau =1/6$ for the $r=0$ special
case only, giving%
\begin{equation}
\sigma (\tau +\Delta \tau )=\frac{1}{6}\left[ 4\sigma (r+\Delta r)+\sigma
(z-\Delta z)+\sigma (z+\Delta z)\right]
\end{equation}%
for $r=0$ pixels. This isn't a perfect solution, but we find it does not
cause significant problems in the code, because it is only applied to a
single row of pixels.

\subsection{The 2D Problem in Hexagonal Coordinates}

We can model planar snow crystal growth using a hexagonal lattice, which can
be mapped onto a rectangular lattice as shown in Figure 1 \cite%
{gravnergriffeath}. The discrete diffusion equation becomes%
\begin{equation}
\sigma (\tau +\Delta \tau )=\frac{2}{3}\Delta \tau
\sum\limits_{i=1}^{6}\sigma _{i}+\left( 1-4\Delta \tau \right) \sigma
\end{equation}%
where the index $i$ refers to the six nearest neighbors around each pixel.
For boundary pixels, we again use this expression and substitute%
\begin{equation}
\sigma _{solid}=\sigma (1-\alpha \Delta \xi )
\end{equation}%
for each neighboring boundary pixel, as above, and we have taken $\Delta x$
to be the distance between nearest neighbors.

We again define an accumulated mass parameter $\lambda $ and
increment it with%
\begin{equation}
\lambda \rightarrow \lambda +\frac{2}{3}\sum \alpha \sigma \Delta \lambda
\end{equation}%
for each time step, where the sum is over the number of ice neighbors. When $%
\lambda \geq 1$, that pixel converts from air to ice.

We label each boundary pixel with $(N),$ where $N$ is the number of nearest
neighbors:

(0) - an air pixel

(1) - a boundary pixel at the tip of a hexagon. The growth will be low here,
and it likely doesn't matter much what value is chosen for $\alpha $. The
growth should be sufficient, however, to allow the crystal to grow from its
initial seed.

(2) - this case refers to the normal growth of a prism facet, and an
analysis shows that we much choose $\alpha =\left( \sqrt{3}/2\right) \alpha
_{prism}$ so the growth equals $v=\alpha _{prism}v_{kin}\sigma $ for this
surface.

(3) - this refers to growth at a kink site, and the choice of $\alpha $ here
has a strong effect on the transition from faceted to branched growth \cite%
{gravnergriffeath}.

(4), (5), (6) - these are unusual cases where the growth is not much
affected by the choice of $\alpha $ as long as it is large (of order unity).

\subsection{The 3D Problem in Hexagonal Coordinates}

The discrete diffusion equation becomes%
\begin{equation}
\sigma (\tau +\Delta \tau )=\frac{2}{3}\Delta \tau
\sum\limits_{i=1}^{6}\sigma _{i}+\Delta \tau \sum\limits_{j=1}^{2}\sigma
_{j}+\left( 1-6\Delta \tau \right) \sigma
\end{equation}%
where the index $i$ refers to the six nearest neighbors around each pixel in
the basal $(xy)$ plane and $j$ refers to the two neighbors perpendicular to
this plane (the $z$ direction). For boundary pixels, we again use this
expression and substitute%
\begin{equation}
\sigma _{solid}=\sigma (1-\alpha \Delta \xi )
\end{equation}%
for each neighboring boundary pixel, as above. We have taken $\Delta x$ to
be the distance between nearest neighbors, which is the same in the basal
plane as in the $z$ direction.

We again define an accumulated mass parameter $\lambda $ and increment it
with%
\begin{equation}
\lambda \rightarrow \lambda +\frac{2}{3}\sum\limits_{i=1}^{6}\alpha \sigma
\Delta \lambda +\sum\limits_{j=1}^{2}\alpha \sigma \Delta \lambda
\end{equation}%
for each time step, where the sums are over ice neighbors. When $\lambda
\geq 1$, that pixel converts from air to ice.

We label each boundary pixel with $(N_{i},N_{j}),$ where $N_{i}$ is the
number of nearest neighbors in the basal plane and $N_{j}$ is the number of
neighbors in the $z$ direction. We will assume that neighbors in the basal
plane are contiguous. Non-contiguous cases are unusual, and it shouldn't
matter much how we assign growth rates in those cases, as long as $\alpha $
is large.

(0,0) - an air pixel

(1,0) - a boundary pixel at the tip of a hexagon, with no $z$ neighbors.
This is similar to the case (1) in the 2D hexagonal problem above.

(2,0) - similar to the (2) case in the 2D hexagonal case above, and we take $%
\alpha =\left( \sqrt{3}/2\right) \alpha _{prism}.$

(3,0) - growth at a kink site, similar to the (3) case above, and again the
value of $\alpha $ chosen will have a strong effect on the transition from
faceted to branched growth \cite{gravnergriffeath}.

(0,1) - growth of the basal facet; so we take $\alpha =\alpha _{basal}.$

(1,1) - not well determined, but faster than (1,0) or (0,1)

(2,1) - not well determined, but faster than (1,1) or (2,0)

(3,1) - again not well determined, but faster than (2,1). The values of $%
\alpha $ used for the $(N_{i},1)$ sites will strongly affect the transition
from faceted growth to structure formation on the basal facet \cite%
{gravnergriffeath}.

The values of $\alpha $ used for the remaining sites should be high, and the
details will likely not affect the growth dynamics significantly.

\section{Using the Algorithms}

\subsection{Intrinsic Anisotropies}

To test our code, we modeled the growth of spherical crystals with isotropic
attachment kinetics, where there is a simple analytic result for the growth
rate. We limited the growth in our models to small changes in radius, to
minimize non-spherical growth that eventually arises from the
Mullins-Sekerka instability \cite{saitobook}. Our numerical model followed
the prescriptions above for the 2D problem in cylindrical coordinates, with $%
\alpha _{x}=\alpha _{y}=\alpha =$ constant for the facet surfaces and $%
\alpha _{11}=\alpha /\sqrt{2}$ for a (1,1) kink site. All higher-order sites
with $B>3$ are irrelevant for this problem, because of the convex geometry
of the spherical surface.

We found that our code generated growth rates that were always a few percent
larger than the analytic theory for all (reasonable) choices of $\alpha ,$ $%
\Delta \xi ,$ $\Lambda ,$ and other input parameters. Upon closer
investigation, we found that our algorithm produces growth of a (12) surface
that is approximately 8 percent faster than the (01), (10), or (11) surfaces
(the latter surfaces all grow at the same rate, which, by design, is the
correct rate).

This demonstrates that our choice of a fixed rectangular grid, along with
only a small number of growth rules, yields an intrinsic anisotropy in the
growth algorithm. This anisotropy could be reduced by adding higher-order
corrections that specify slightly different $\alpha $ values for different
boundary pixels, depending on the surface configuration beyond just the
nearest neighbors, but such corrections would be difficult to implement.

This intrinsic anisotropy should not be a significant effect for strongly
faceted growth, but we expect that our LCA approach would not produce
accurate morphological results for cases where the anisotropy in the
attachment kinetics is less than approximately 10 percent.

\subsection{Scaling Behavior}

If we run a growth code and produce some complex crystal shape, the
interpretation of our result still contains an ambiguity. The crystal size
is given in pixels, where $\Delta x=\Delta \xi X_{0}$ is the pixel size. The
parameter $\Delta \xi $ was fixed in the code, but $X_{0}$ depends on the
diffusion constant $D$, which is not otherwise specified. Similarly, a
single time step in the code corresponds to a physical time%
\begin{eqnarray}
\Delta t &=&\frac{\left( \Delta x\right) ^{2}}{D}\Delta \tau \\
&=&\frac{X_{0}^{2}\Delta \xi ^{2}\Delta \tau }{D}\sim D  \nonumber
\end{eqnarray}

Thus we see that the growth behavior at different air pressures (different $%
D)$ is determined once we know the growth at a single pressure (provided $%
\sigma _{\infty }$ is the same at the different pressures). If the air
pressure is half an atmosphere, the growth morphology (however complex) will
be the same as at one atmosphere, except in the former case the crystal will
be double the size in double the time. This scaling behavior nicely explains
why snow crystal morphology is generally simpler for smaller crystals and/or
for lower air pressures, which has long been observed \cite{libbrechtreview}%
. To my knowledge, this scaling behavior has not been identified in previous
investigations of snow crystal growth.

\subsection{Limitations on the Grid Size}

The physical size of the grid is $\Delta x=\Delta \xi X_{0},$ and there are
limits to how coarse the grid can be without affecting the growth or causing
instabilities in the code. Taking $\Delta \xi >1/\alpha $ would cause $%
\sigma _{solid}$ to become negative, which causes some concern in that it
may produce instabilities in the code. With this limitation, the grid
spacing could not be larger than $\Delta x=X_{0}/\alpha .$ For air at a
pressure of one atmosphere and $\alpha \approx 1,$ this gives $\Delta x=$
0.15 $\mu $m. Modeling a 1.5 mm snow crystal would then require a grid with
at least 10,000 pixels on a side, which is something of a computational
challenge.

We can do better by realizing that a negative $\sigma _{solid}$ is not
itself sufficient to cause instabilities. A closer look at the algorithm
reveals that problems only begin when $\sigma (\tau +\Delta \tau )$ becomes
negative in a single timestep, which happens when the supersaturation is
"drained" to a negative value according to Equation \ref{changingsigma}.
This puts a limitation of $\Delta \xi >1/\alpha \Delta \tau $ on the grid
size, so we are able to use a coarser grid spacing if we also use a finer
time step. Whatever the limit, it is important to note that our choice of
grid spacing is not simply limited by a desire to produce a physically
accurate model of crystal growth, but also by intrinsic instabilities in the
code.

Physically, we can gain some insights into these limitations from dendrite
growth theory \cite{libbrechtreview}. We have $X_{0}\approx R_{kin}$ (the
latter from Equation 28 in \cite{libbrechtreview}), and a growing dendrite
has a tip radius%
\begin{eqnarray}
R_{tip} &=&\frac{B}{\alpha s}R_{kin} \\
&\approx &\frac{X_{0}}{\alpha s}  \nonumber
\end{eqnarray}%
where $s$ is the dimensionless solvability parameter, which is of order
unity for ice crystal growth \cite{libbrechtreview}. The stability of the
code (with $\Delta \tau \approx 1)$ thus limits the grid spacing to be no
greater than the tip radius of a growing dendritic structure. Interestingly,
it appears that the code can only function properly when the grid spacing is
fine enough to allow the growth of physically realistic dendritic
structures, the scale of which is given by solvability theory.

\subsection{Adaptive Time Steps}

We wish to choose $\Lambda $ in Equation \ref{lambda} as large as possible,
so the code runs quickly, while not altering the growth appreciably. Our
first criterion is that the growth be slow over a single time step, so that
it takes at least $N_{0}\approx 10$ time steps before a boundary pixel turns
to ice, which means we must take%
\begin{equation}
\Lambda <\frac{1}{\alpha \sigma \Delta \xi \Delta \tau N_{0}}
\end{equation}%
Another criterion is that the Peclet number should be small, as dictated by
the physics of the growth problem, for which we find%
\begin{equation}
\Lambda <\left( \frac{\Delta x}{R}\right) \frac{\Delta \tau }{\Delta \xi }%
\frac{1}{\alpha \sigma }
\end{equation}%
where $R$ is a characteristic size of the crystal. Since $R_{i}=R/\Delta x$
is typically larger than $N_{0},$ the latter requirement is the more
stringent of the two.

We can speed up the code further by taking 
\begin{equation}
\Lambda =A\frac{1}{R_{i,\max }}\frac{1}{\left( \alpha \sigma \right) _{\max }%
}
\end{equation}%
where $A<1$ is a constant, $R_{i,\max }$ is the current maximum size of the
crystal (in pixels), and $\left( \alpha \sigma \right) _{\max }$ is the
maximum product of $\alpha $ and $\sigma $ over all current boundary pixels.
This speeds up the code considerably when the growth is strongly diffusion
limited (so $\sigma \ll \sigma _{\infty })$ or when $\alpha $ is small on
the crystal surface.

This choice of $\Lambda $ means essentially using an adaptive time step,
where the physical time for each step is equal to%
\begin{equation}
\Delta t=\Lambda \Delta \xi ^{2}\Delta \tau \Delta t_{0}
\end{equation}%
where%
\begin{eqnarray}
\Delta t_{0} &=&\frac{2\pi m}{kT}D\frac{c_{solid}}{c_{sat}} \\
&\approx &\left( \frac{D}{D_{air}}\right) \times (1 \  msec) \nonumber
\end{eqnarray}%
and the latter value is for growth at $T=-15$ C \cite{libbrechtreview}.

\section{A Morphological Transition in Plate-like Growth}

We have applied the algorithms derived above to examine a key problem in
snow crystal growth (and, by extension, other examples of faceted crystal
growth) - understanding how small changes in extrinsic parameters like
temperature and supersaturation can produce rather dramatic changes in the
resulting crystal morphologies. Using our new LCA growth algorithms, we
discovered an interesting example of a morphological transition in the
growth of plate-like snow crystals.

We modeled the growth of small snow crystal plates in the limit of
cylindrical symmetry, so we could use the 2D algorithm in cylindrical
coordinates described above. This is much simpler computationally than the
full 3D problem, and the approximation is a reasonable one for small plates
without dendritic branching \cite{kglmodel}. Instead of six prism facets,
the cylindrical model produces one continuous "facet" which is the perimeter
of the plate. Thus, for example, hollowing of the six prism faces is
replaced by hollowing of the perimeter "facet". Aside from geometrical
factors of order unity, we believe the cylindrical model is a good
approximation for the growth of simple plate-like snow crystals.

In our model we used a grid of $N_{r}=200$ by $N_{z}=100$ pixels with $%
\Delta \xi =1,$ corresponding to a physical grid spacing of $\Delta x=0.15$ $%
\mu $m. The boundary condition on the $z=0$ plane guaranteed symmetry about
that plane. We began each growth run with a single ice pixel at $r=z=0,$ and
we ran the code until the maximum crystal radius reached 20 $\mu $m. For the
basal surface, we assumed that the growth was nucleation-limited with $%
\alpha (\sigma )=\min [1,A\exp (-\sigma _{0}/\sigma )]$ with $A=2$ and $%
\sigma =0.021,$ following the latest ice crystal growth measurements \cite%
{libbrechtnew}. Accurate prism growth measurements do not yet exist, so we
assumed nucleation-limited growth with $A=5$ and $\sigma _{0}=0.01.$ For
non-facet surfaces we chose $\alpha _{11}=0.7$ for the kink site with $B=3$
and $\alpha =1$ for all sites with $B>3.$ We believe that the morphological
transition we observed is insensitive to the exact values of these growth
parameters, as long as they produce plate-like structures with
nucleation-limited growth on both facets.

Our results are shown in Figures 2 and 3 as we varied only the
supersaturation $\sigma _{\infty }$ in the model. The figures show a clear
morphological transition from simple, thin-plate growth at low $\sigma
_{\infty }$ to concave (hollowed) growth at intermediate $\sigma _{\infty }$
and convex growth at high $\sigma _{\infty }.$

The growth of thin plates at low $\sigma _{\infty }$ (number 1 in the
figures) is easy to understand from simple considerations of the growth
dynamics. At low $\sigma _{\infty },$ $\alpha $ is small and the growth is
largely limited by attachment kinetics, so $\sigma \approx $ $\sigma
_{\infty }$ at the crystal surface. The relative growth velocities of the
prism and basal faces is then%
\begin{equation}
\frac{v_{prism}}{v_{basal}}\approx \frac{A_{prism}}{A_{basal}}\exp \left[
\left( \sigma _{0,basal}-\sigma _{0,prism}\right) /\sigma _{\infty }\right]
\end{equation}%
Since $\sigma _{0,basal}>\sigma _{0,prism}$ for our case, this ratio
increases rapidly as $\sigma _{\infty }$ decreases, producing thinner plates
at lower $\sigma _{\infty }.$

As $\sigma _{\infty }$ increases, the higher $\sigma $ values at the corners
causes hollowing of both the prism and basal facets, as seen in Figure 3,
number 2. For this concave growth, we see that steps are generated at the
edges of the basal facets and propagate inward as the crystal grows.

As $\sigma _{\infty }$ increases still further, the growth undergoes a
rather sudden transition to convex growth, seen in Figure 3, number 3. This
transition is accompanied by a reduction in crystal thickness and growth
time, as seen in Figure 2. These two are related in that a thinner crystal
requires less mass for a given radius, and thus grows to a given radius (20 $%
\mu $m in this case) in less time. For this convex growth, steps are
generated at the centers of the basal facets and propagate outward, in
contrast to the concave case.

At the highest $\sigma _{\infty }$ shown, the basal growth has increased
until additional structure is seen at the center of the basal facet. Similar
structures appear when the growth rates of the prism and basal facets are
comparable, and this growth behavior reflects the fact that $\alpha
_{basal}/\alpha _{prism}$ increases with increasing $\sigma _{\infty }$ as
the surface values of $\sigma $ begin to exceed the critical
supersaturations $\sigma _{0}$ for the facets.

The transition from concave to convex growth as the supersaturation
increases is well-known in snow crystal growth \cite{libbrechtreview},
because it is one of the most studied examples of faceted crystal growth.
Other crystal systems may exhibit similar morphological transitions as the
growth drive is increased. The morphologies that result certainly depend on
the attachment kinetics, so the detailed modeling must be tailored to each
individual system.

The LCA algorithms described above have proven to be quite robust and
numerically stable in these calculations, while being simple to code with
reasonably fast execution times. It is straightforward to change the
attachment kinetics and other growth parameters to investigate different
regimes. The LCA method is particular well-suited to modeling the horizontal
propagation of macrosteps on a flat surface, which appears to be a key
element of faceted growth. Further studies are needed to identify whatever
numerical idiosyncrasies are inherent in the LCA method (such as the
intrinsic anisotropy described above), but so far it appears to be a
powerful and flexible approach for modeling faceted crystal growth.

\begin{figure}[tbp] 
  \centering
  \includegraphics[width=4.5in,keepaspectratio]{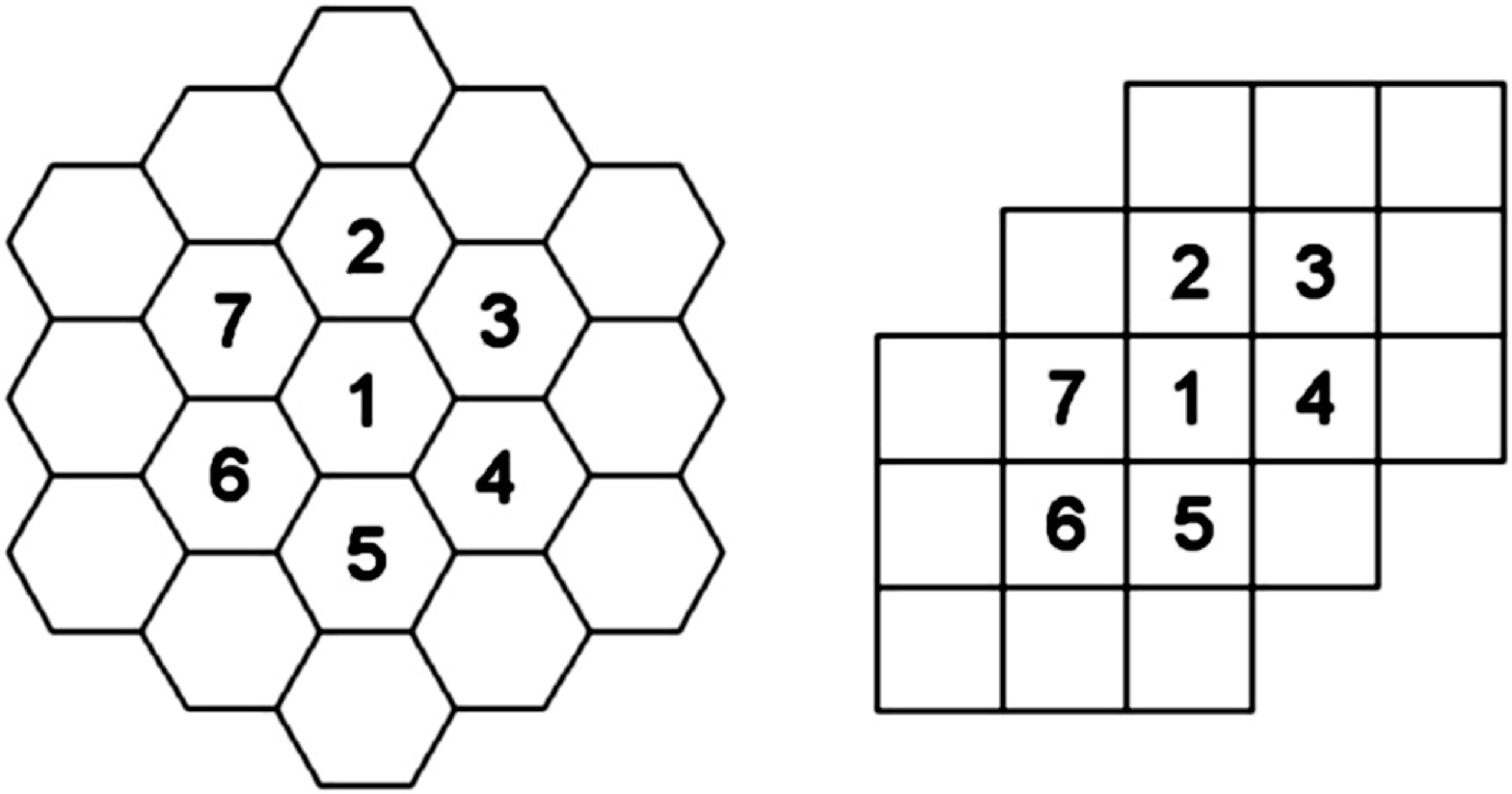}
  \caption{A mapping of a hexagonal grid onto a rectangular grid. Corresponding pixels are numbered, showing the arrangement of nearest neighbors in each case.}
  \label{fig:hexnotation2}
\end{figure}

\begin{figure}[tbp] 
  \centering
  \includegraphics[width=4.5in, keepaspectratio]{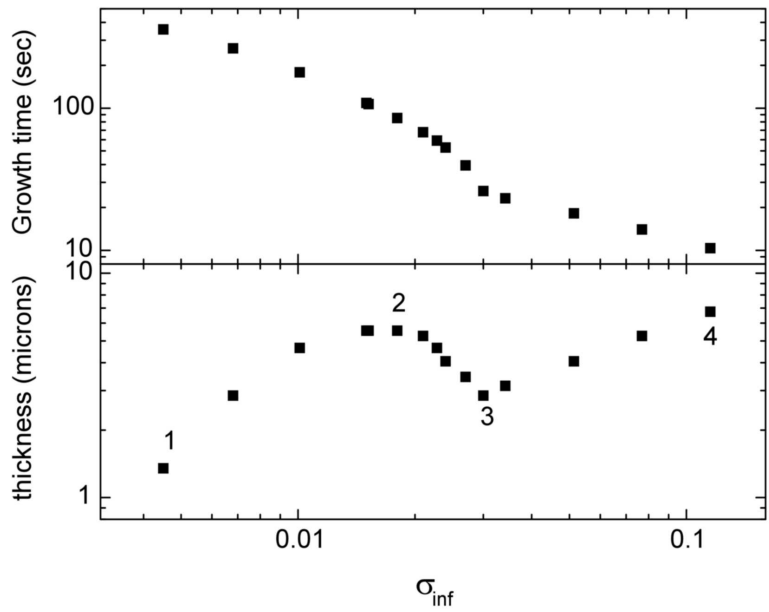}
  \caption{Modeled growth time (top panel) and final thickness (lower panel) of a crystal as a function of the supersaturation at the outer boundary. For all crystals the growth was terminated when the diameter reached 40 $\mu $m. Growth parameters are described in the text. The numbers in the lower panel correspond to those in Figure 3 showing the crystal morphology. }
  \label{fig:diskgrowth}
\end{figure}

\begin{figure}[tbp] 
  \centering
  \includegraphics[width=4.5in,keepaspectratio]{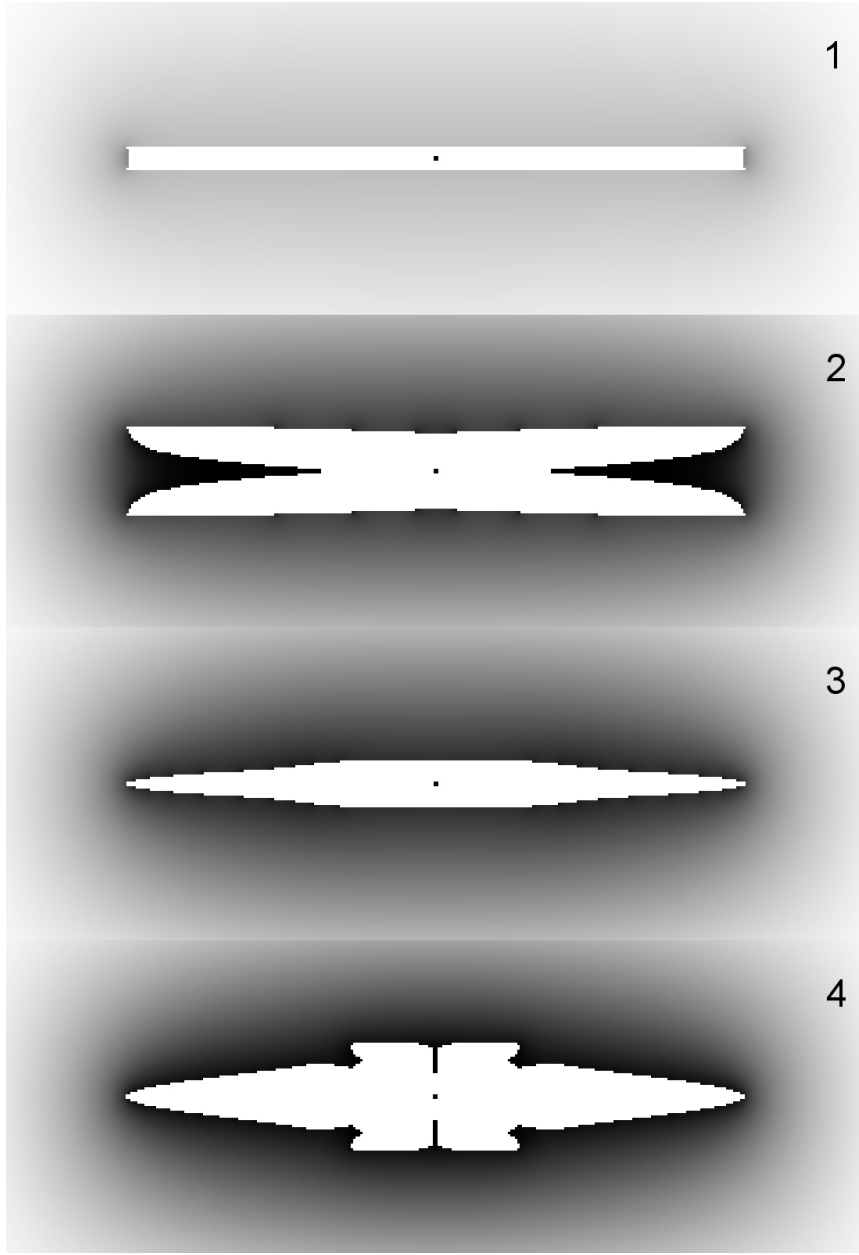}
  \caption{Full cross-sections of four plate-like crystals at the end of their growth, at which time the diameter was 40 $\mu $m. The numbers beside each crystal correspond to those in Figure 2. This shows the transition from simple plate growth (1) to concave growth (2) to convex growth (3 and 4) as $\sigma _{\infty}$ is increased.}
  \label{fig:combined}
\end{figure}

\end{document}